\documentclass[prl,showpacs,twocolumn]{revtex4}
\usepackage{amssymb}
\usepackage{amsmath}
\usepackage{graphicx, ulem}

\usepackage{color,colordvi}
\definecolor{blue}{rgb}{0,0,1}
\definecolor{grey}{rgb}{0.6,0.6,0.6}

\def \be{\begin{equation}}
\def \ee{\end{equation}}
\def \bew{\begin{widetext}\begin{equation}}
\def \eew{\end{equation}\end{widetext}}
\def \bmlett{\begin{mathletters}}
\def \emlett{\end{mathletters}}
\def    \bse{\begin{subequations}}
\def    \ese{\end{subequations}}
\def\bee{\begin{eqnarray}}
\def\eee{\end{eqnarray}}

\def\kb{k_{\rm B}}

\def\halb{\mbox{$\frac{1}{2}$}}

\newcommand{\bbbone}{{\mathchoice {\rm 1\mskip -4mu l}{\rm 1\mskip -4mu l}{\rm 1\mskip -4.5mu l}{\rm 1\mskip -5mu l}}}

\def \kbT{k_{\rm B} T}
\def \hH{\hat{H}}
\def \hc{\hat{c}}
\def \hsigma{\hat{\sigma}}

\def \xiN{\xi_{\rm N}}

\begin{document}

\title{Quantum Kibble-Zurek physics in the presence of spatially-correlated dissipation}
\author{P. Nalbach$^{1,2}$, Smitha Vishveshwara$^3$, and Aashish A. Clerk$^4$}
\affiliation{
$^1$I.\ Institut f\"ur Theoretische Physik,  Universit\"at
Hamburg, Jungiusstra{\ss}e 9, 20355 Hamburg, Germany\\
$^2$The Hamburg Centre for Ultrafast Imaging, Luruper Chaussee 149, 22761 Hamburg, Germany\\
$^3$Department of Physics, University of Illinois at Urbana-Champaign, Urbana, Illinois 61801-3080, USA\\
$^4$Department of Physics, McGill University, 3600 rue
University, Montreal, QC Canada H3A 2T8}

\date{\today}

\begin{abstract}

We study how  universal properties of  quantum quenches across critical points are modified by a weak coupling to thermal dissipation, focusing on the paradigmatic case of the transverse field Ising model. Beyond the standard quench-induced Kibble-Zurek defect production in the absence of the bath, the bath contributes extra thermal defects. 
We show that spatial correlations in the noise produced by the bath can play a crucial role:  one obtains quantitatively different scaling regimes depending on whether the correlation length of the noise is smaller or larger than the Kibble-Zurek length associated with the quench speed, and the thermal length set by temperature.  For the case of spatially-correlated bath noise, additional thermal defect generation is restricted to a window that is both quantum critical and excluded from the non-equilibrium regime surrounding the critical point. We map the dissipative quench problem to a set of effectively independent dissipative Landau-Zener problems.  Using this mapping along with both analytic and numerical calculations allows us to find the scaling of the excess defect density produced in the quench, and suggests a generic picture for such dissipative quenches.
\end{abstract}


\maketitle

% 64.60.Ht 	Dynamic critical phenomena 
% 64.70.Tg 	Quantum phase transitions
% 05.30.Rt 	Quantum phase transitions (see also 64.70.Tg Quantum phase transitions in specific phase transitions; and 73.43.Nq Quantum phase transitions in Quantum Hall effects) 
% 03.65.Yz 	Decoherence; open systems; quantum statistical methods (see also 03.67.Pp in quantum information; for decoherence in Bose-Einstein condensates, see 03.75.Gg)

%%%%%%%%%%%%%%%%%%%%%%%%%%%%%%%%%%%%%%%%%%%%%%%%%%%%%%%%%%%%%%%%%%%%%%%%%
%%%%%%%%%%%%%%%%%%%%%%%%%%%%%%%%%%%%%%%%%%%%%%%%%%%%%%%%%%%%%%%%%%%%%%%%%

\textit{Introduction-- } 
Quantum quenches involve the explicit time-dependent tuning of a Hamiltonian and are among the most basic and generic of phenomena in quantum many-body dynamics. They have garnered considerable recent interest,  particularly with regards to tuning through a quantum critical point, for which the quantum Kibble-Zurek (KZ) mechanism forms a standard paradigm~\cite{PolkovnikovRMP, Dziarmaga2005, Polkovnikov2005, Zurek2005, Dziarmaga2010, Dutta2010}, analogous to the corresponding argument for classical, thermal quenches~\cite{Kibble1976,Kibble1980,Zurek1985,Zurek1996}.  The basic idea is that for gapped quantum phases separated by a gapless critical point, as one approaches the critical point, the system's relaxation time (i.e.~inverse gap, $\Delta^{-1}$) diverges.  When this relaxation time becomes comparable to the time associated with the quench speed $v$, the system falls out of equilibrium. This non-equilibrium regime dictates post-quench behavior, such as deviations from the final ground state and corresponding defect densities.  One obtains universal scaling behavior of such quantities, ultimately governed by the proximity to the quantum critical point. 

In this work we turn our attention to the important yet relatively unstudied effect of thermal dissipation on the quantum Kibble-Zurek scenario. We focus on the prototypical system of a quench in a transverse-field Ising model (TFIM), where a dissipative thermal bath produces noise in the transverse field on each lattice site. Our study reveals that such dissipative quantum quenches are strongly influenced by the interplay between different fundamental length scales. 

A crucial role is played by the ratio of the bath-noise correlation length $\xi_N$ to both the length scale of the dissipation-free Kibble Zurek problem 
 $\xi_{\rm KZ}\propto 1/\sqrt{v}$ and to the thermal length scale $\xi_{\rm T}\propto 1/\kb T$.  
Patan\`e et al.~\cite{Patane2008, Patane2009} investigated the limit $\xiN \ll \xi_{\rm KZ}, \xi_{\rm T}$, where spatial noise correlations are essentially irrelevant, and each lattice site is effectively coupled to an independent bath.  In contrast, we analyze the opposite but equally important limit where the noise correlation length, while finite, is nonetheless larger than both $\xi_{\rm KZ}$ and $\xi_{\rm T}$.  In this case, the long-range bath noise correlations lead to strikingly different behaviours.  

\begin{figure}
\includegraphics[width=0.35\textwidth]{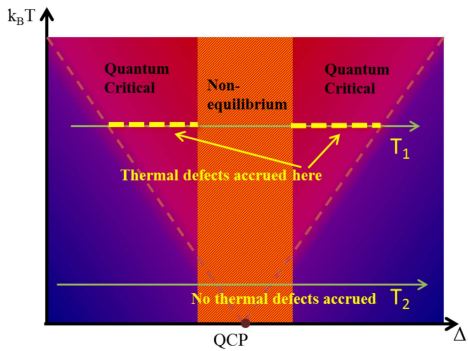}
\caption{(Color online) Quench trajectories for two temperatures $T_1$ and $T_2$ at a fixed quench rate. At $T_1$, far from the quantum critical point (QCP), the spectral gap $\Delta$ is too large for thermal excitations to be created. Thermal defect generation thus turns on only once the trajectory enters the quantum critical regime $k_BT>\Delta$.  It is however suppressed again once the system enters the non-equilibrium regime near the QCP, where the intrinsic relaxational rate is slower than the quench rate.  The thick yellow dashed line indicates the portion of the quench where thermal defects are produced.   For low temperature ($T_2$), the entire quantum critical regime is subsumed within the non-equilibrium regime near the QCP, and hence there is no appreciable thermal defect generation.  }
\label{fig:dissquench}
\end{figure}

For this correlated noise regime and generic form of system-bath coupling, our work suggests the following general picture, as graphically depicted in  Fig.~\ref{fig:dissquench}. 
During the quench, thermal effects come into play only under certain conditions.
First, at a given time $t$, the bath temperature needs to be large enough to be able to produce defects, implying that one must be in the quantum critical regime $k_BT > \Delta(t)$~\cite{Sachdev}. 
 However, this is not enough:  even if the temperature is large enough, the coherent system dynamics must be able to equilibrate in order for the weak dissipation to also play a role.  This means that thermal defect generation is suppressed in the ``non-equilibrium" Kibble-Zurek  regime close to the critical point.  
Thus,  dissipation only gives rise to additional defects during (at most) a limited portion of the quench protocol.

%%%%%%%%%%%%%%%%%%%%%%%%%%%%%%%%%%%%%%%%
%%%%%%%%%%%%%%%%%%%%%%%%%%%%%%%%%%%%%%%%
 
Focusing on temperatures small enough that only long-wavelength excitations can be produced, the above arguments lead to a general prediction that a weakly-coupled  bath only gives rise to additional defect generation (above the standard zero-temperature Kibble-Zurek prediction) if the temperature satisfies $k_B T> k_B T_{\rm min}$. As detailed in the EPAPS \cite{suppInfo},we find that the Kibble-Zurek criterion for the cross-over scale into the non-equilibrium regime~\cite{PolkovnikovRMP, Chandran2012, Dziarmaga2010} combined with constraints for being in the quantum critical regime result in the following scaling behavior:
\begin{equation}
%	(k_BT)^{(1+\nu z)/(\nu z)}>v ,
	k_B T_{\rm min} \propto v^{\nu z / (1+\nu z)} .
\end{equation}
Here $\nu$ is the critical exponent describing the divergence of the system correlation length near the transition, and $z$ is the corresponding dynamical critical exponent. For the specific case of the transverse field Ising model (where $\nu = z = 1$), we find that the excess defect density scales as $(\kb T)^3/v$ as long as $\sqrt{v}\ll \kb T$, but is strongly suppressed for lower temperatures.

To rigorously establish the above picture, we utilize a major technical advantage of large $\xiN$:  one can map the dissipative quench problem to an ensemble of {\it dissipative} Landau-Zener problems~\cite{Landau1932, Zener1932, Stueckelberg1932, Majorana1932}.  We justify this mapping in what follows, and then make use of both 
analytic and numerical approaches to calculate the production of defects in the quench in the presence of a weak coupling to spatially-correlated dissipation. 

%%%%%%%%%%%%%%%%%%%%%%%%%%%%%%%%%%%%%%%%%%%

\textit{Model-- }  We consider a one-dimensional TFIM subject to noise in the transverse field via a coupling to a thermal bosonic bath. 
We primarily focus on the limit where the bath noise is spatially uniform, like the average transverse field itself.
The net Hamiltonian takes the form $\hH = \hH_S + \hH_B + \hH_{SB}$, where
\begin{equation}
	 \hH_{S} + \hH_{SB} = -J \sum_{j} \hsigma^x_j \hsigma^x_{j+1} - \left(h + \hat{X} \right) \sum_j \hsigma_j^z
\label{eq:HamTI}
\end{equation}
represents the TFIM system and system-bath coupling Hamiltonian, $H_B = \sum_\nu \omega_\nu \hat{b}^\dag_\nu \hat{b}_\nu$ is the bath Hamiltonian and
$\hat{X} = \sum_\nu \lambda_\nu \left( \hat{b}_\nu + h.c. \right)$.
Here, $\hsigma^{x,z}_j$ denote Pauli matrices for the spin at site $j$, $J$ is the exchange coupling, $h$ a Zeeman field in the $z$-direction, and we have set $\hbar=1$.
Without loss of generality, we take $J, h \ge 0$.  The system is in an Ising ferromagnetically ordered
phase when $h < J$, while for $h>J$ the system is in a paramagnetic phase. The two phases are separated by a quantum critical point at $h_c=J$.
The bath is characterized by the spectral density
 $\mathcal{J}(\omega)= \sum_\nu \lambda_\nu^2 \delta(\omega-\omega_\nu) = \gamma \omega^s \omega_c^{1-s} e^{-\omega/\omega_c}$, where
$\gamma$ is the dimensionless coupling strength and $\omega_c$ is a cut-off frequency.  We focus on the standard case of an Ohmic bath, where $s=1$.

Employing the standard Jordan-Wigner transformation \cite{JordanWigner}, we re-express the Hamiltonian in terms of spinless fermions, $\hc_i = ( \prod_{j<i} \hsigma_j^z ) ~\hsigma_i^+$.
Working in momentum space and setting the lattice constant $a=1$, we have:
\begin{eqnarray}
	 \hH_S + \hH_{SB} & = &
	 	\sum_{0 \le k \le \pi} \left[\begin{array}{cc}
			\hc^\dagger_{k} & \hc_{-k}
		\end{array} \right]
		\left( \mathcal{H}_{k,S} + \mathcal{H}_{k,SB} \right)
\left[\begin{array}{c}
			\hc_{k}
			\\ \hc^\dagger_{-k}
	\end{array} \right]  \nonumber \\
	\label{eq:Hamf}		
&& \hspace*{-2.3cm}{\rm with}\quad 
			 \mathcal{H}_{k,S}  =
		\left[\begin{array}{cc}
			\xi_k & \Delta_k \\ \Delta_k^* & -\xi_k \end{array} \right],
			\quad
		 \mathcal{H}_{k,SB}  =
		\left[\begin{array}{cc}
			\hat{X} & 0 \\ 0 & -\hat{X}  \end{array} \right],
\end{eqnarray}
and where $\xi_k = 2h -2 J \cos (k)$ and $\Delta_k = 2 J \sin(k)$.
The corresponding energy dispersion of $\hH_S$ is given by  $\epsilon_k = \sqrt{\xi_k^2 +|\Delta_k|^2}$.
Critical exponents $\nu=z=1$ can be extracted from critical gap behavior, namely,
$\Delta_{k=0} \sim |h-h_c|$ (near $h = h_c=J$) and $\Delta_k \sim |k|$ (for $k \rightarrow 0$ when  $h=h_c$).

\textit{Mapping to the Landau-Zener problem-- }
We briefly recall the standard treatment of quench dynamics in the absence of dissipation~\cite{PolkovnikovRMP}.  Each term in $\hH_S$ (c.f.~Eq.~(\ref{eq:Hamf})) describes an effective two-level system, where the two states correspond to having the orbitals $(k,-k)$ either both empty or both occupied. The most common quench protocol involves a linear ramp of the form
\begin{equation}
%	h(t)=h_c+vt.
	h(t)= -vt.
\label{eq:quench}
\end{equation}
For concreteness, we consider a quench that starts in a paramagnetic phase at $t = -\infty$ and ends in the ferromagnetic phase at $t=0$. With this time dependence, each term in $\hH_S$ describes a Landau-Zener problem ~\cite{Landau1932, Zener1932, Stueckelberg1932, Majorana1932}:  a two level system subject to a constant $x$ magnetic field ($\Delta_k$) and linearly time-varying $z$ magnetic field ($\xi_k = -2 v t - 2 J \cos k$), which is taken through the avoided crossing occurring at $\xi_k = 0$.

Evaluating the density of defects $n_{\rm KZ}$ produced by the quench now amounts to calculating the final population of quasiparticles at the end of the quench.
For slow quenches  (with velocity $v \ll J^2 $), the dominant contribution to this population comes from low-$k$ fermion modes, where the gap $\Delta_k$
at the avoided crossing is the smallest.  As discussed in
Ref.~\cite{Dziarmaga2005}, the excitation probability of such modes is well-described by the asymptotic, infinite-time LZ probability that the two-level system (TLS) transitions to the final excited state~\cite{Landau1932, Zener1932, Stueckelberg1932, Majorana1932}:
\begin{equation}
	P_{k} \sim \exp{(-\pi \Delta_k^2/  v)}
	\equiv \exp \left(- \pi z_k \right),
	\label{eq:LZProb}
\end{equation}
where we have introduced the adiabaticity parameter $z_k=\Delta_k^2/v$.  Using this expression and integrating over all momentum modes, one obtains the power-law form $n_{\rm KZ} \sim \sqrt{v}$, consistent with the Kibble-Zurek form \cite{Dziarmaga2005}.

%%%%%%%%%%%%%%%%%%%%%%%%%%%%%%%%%%%%%%%%%%%
\textit{Mapping to dissipative Landau-Zener problem-- }
Upon including the uniform coupling to the bath,
we see from Eq.~(\ref{eq:Hamf}) that the translational invariance of the fermionic system is maintained. Thus, for each $k>0$, we have an effective two-level system whose detuning is linearly ramped in time and fluctuates now. The external noise $\hat{X}$, however, couples to every $k$ mode. The single-particle fermion Green functions for each momentum $k$, however, are decoupled from one another (see \cite{suppInfo} for details). As a result, the defect density can be rigorously calculated by assuming that each $k$ mode is coupled to its own, independent dissipative bath. 
This makes the mapping of our dissipative TFIM onto an ensemble of dissipative Landau Zener systems complete for the quantity of interest.
Such dissipative LZ problems with diagonal noise have been well-studied in the literature \cite{Ao1989,Ao1991,  Shimshoni1991, Kayanuma1998, Wubs2006, NalbachLZ09,NalbachLZ10, Orth2010, Orth2013}. 

%%%%%%%%%%%%%%%%%%%%%%%%%%%%%%%%%%%%%%%%%%%
\textit{Analytic estimates-- }
The above mapping enables us to invoke results for the dissipative LZ problem, first established by  Ao and Rammer~\cite{Ao1989,Ao1991}, to evaluate the contribution of thermal dissipation towards defect production during the quench. 
For modes having adiabaticity parameter $z_k \lesssim 1$ (c.f.~Eq.~(\ref{eq:LZProb})),
the transition through the avoided crossing is not adiabatic,
and there is a large probability for the effective TLS to be excited even without dissipation.  For such modes, dissipation (to leading order) {\it does not} yield any additional probability of excitation beyond the LZ expression in Eq.~(\ref{eq:LZProb}).  Note that dissipation can only give rise to transitions between eigenstates which are coherent superpositions of $\sigma_z$ eigenstates.  While
the instantaneous eigenstates have this form near the avoided crossing, for the fast modes with $z_k \lesssim 1$, there is not enough time for these eigenstates to form.  

In contrast, for modes having $z_k > 1$, the transition through the avoided crossing is very nearly adiabatic; without dissipation, they remain close to the ground state.  For such ``slow" modes, the system spends enough time near the avoided crossing for dissipation to give rise to transitions to the excited state.  This leads to an additional dissipation-induced excitation probability $\delta P_k$ that for weak dissipation simply {\it adds} to the LZ expression in  Eq.~(\ref{eq:LZProb}).  The probability $\delta P_k$ was explicitly calculated in Refs.~\cite{Ao1989,Ao1991} and has a simple form consistent with a Golden rule calculation:
\begin{equation}
	\delta P_k = \frac{2 \pi \Delta_k}{v} \mathcal{J}(2\Delta_k) \, n_B[ 2\Delta_k, T],
\label{eq:ProbTherm}
\end{equation}
where $n_B[E,T]$ is a Bose-Einstein distribution evaluated at energy $E$ and at the temperature $T$ of the dissipative bath.  This is simply the Fermi's Golden Rule rate for excitation of the TLS at the avoided crossing via absorption of a bath phonon ($\Gamma_{\rm exc} = 2 \pi\mathcal{J}(2 \Delta_k) n_B(2 \Delta_k)$), multiplied by the effective time spent at the avoided crossing ($t_{\rm cross} \sim \Delta_k / v$).

Hence, as sketched in Fig.~\ref{fig:dissquench}, the dissipation only influences modes where $\Delta_k$ is sufficiently large to yield a near-adiabatic transition, but also small enough that the Bose-Einstein factor in Eq.~(\ref{eq:ProbTherm}) is appreciable (i.e.~small enough that the bath temperature is sufficient to generate an excitation).  Thus, the only modes affected by dissipation satisfy:
\begin{equation}
	\left( \Delta_{KZ} = \sqrt{v} \right) \lesssim \Delta_k \lesssim \kbT.
	\label{eq:DissRange}
\end{equation}
For sufficiently slow quench velocities, all the relevant modes satisfying this condition correspond to small $k$, where $\Delta_k \sim J k$.  This justifies the general picture provided in Fig.~\ref{fig:dissquench}. Outside the quantum critical regime, $\Delta \equiv \Delta_{k=0} > T$, and there are no modes which satisfy Eq.~(\ref{eq:DissRange}); dissipation thus has no effect here (i.e.~blue/purple region in Fig.~\ref{fig:dissquench}). In the non-equilibrium regime (orange in Fig.~\ref{fig:dissquench}), $\Delta_{k=0} < \Delta_{KZ}$ and v is too large for dissipation to affect any of the modes. Finally, in the portion of the quantum-critical region that lies outside the non-equilibrium regime, 
we have instead
$\Delta_{KZ} < \Delta_{k=0} < \kbT$; we thus have modes satisfying Eq.~(\ref{eq:DissRange}) and dissipation can create excitations. Here, we observe dynamics different from the KZ scenario.

Integrating $\delta P_k$ of Eq.(\ref{eq:ProbTherm}) over the range defined by Eq.~(\ref{eq:DissRange}), and assuming $\kbT \gg \sqrt{ v}$,
we find that the total dissipation-induced excitation density obeys:
\begin{equation}
	n_{\rm th}\sim\lambda^2 k_BT[ (k_BT)^2/v].
\label{eq:nth}
\end{equation}
This result is consistent with a more general ansatz for a scaling form for the dissipation-induced defect density, see EPAPS \cite{suppInfo}.

%%%%%%%%%%%%%%%%%%%%%%%%%%%%%%%%%%%%%%%%%%%
\textit{Numerics --} To confirm the picture presented above, we perform a rigorous numerical analysis of a TFIM quench within the dissipative LZ framework. We employ weak system-bath coupling techniques where simple but approximate Markovian approaches, specifically nonequilibrium Bloch equations (NBEs) \cite{NalbachLZ13}, have been shown to be reliable \cite{NalbachLZ14} in comparison to numerically-exact approaches \cite{NalbachLZ09,NalbachLZ10} (details in \cite{suppInfo}).

%%%%%%%%%%%%%%%%%%%%%%%%%%%%%%%%%%%%%%%%
\begin{figure}
\includegraphics[width=0.47\textwidth]{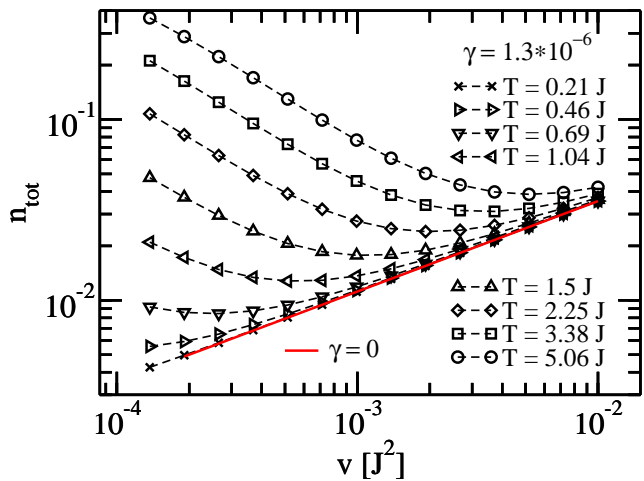}
\caption{(Color online) Total excitation density $n_{\rm tot}(v;\gamma,T)$ versus quench speed $v$ for various temperatures, $\omega_c\gg J$, and a system-bath coupling $\gamma=1.3\cdot 10^{-6}$. The red full line shows $n_{\rm KZ}(v)$, i.e. the excitation density for $\gamma=0$.}
\label{fig:NumRes1}
\end{figure}

%%%%%%%%%%%%%%%%%%%%%%%%

The quench is explicitly started at $h(t_0) = -8 J$ and ended at $h(0)=0$, implying $v = -8 J / t_0$.
We have ensured that results do not depend on this initialization. Solving the adiabatic-Markovian master equation for each $k$, we obtain the total probability $P(k,v;\gamma,T)$ for a given effective TLS to end in the excited state. The thermal contribution $P_{\rm th}(k,v;\gamma,T)$ is determined by subtracting the dissipation-free probability $P(k,v;\gamma=0,T)$ from the total one. Integrating over all $k$ we find the total excitation density produced during the quench,  
\[
n_{\rm tot}(v;\gamma,T) = \int_0^\pi dk \; P(k,v;\gamma,T)
\equiv n_{\rm KZ}(v) + n_{\rm th}(v; \gamma,T),
\]
where $n_{\rm KZ}$ is the defect density without dissipation, and $n_{\rm th}$ is the additional defect density due to the thermal dissipation.

In Fig.\ \ref{fig:NumRes1} we plot the total defect density produced by the quench as a function of quench velocity $v$; different curves correspond to different bath temperatures.  The defect density for the dissipation-free case $\gamma=0$ is also plotted (red solid line); this curve exhibits the standard $\sqrt{v}$ Kibble Zurek scaling. For finite $\gamma$ and $T$ we observe with decreasing $v$ an increasing defect probability due to additional thermal defects. Thus, at a fixed temperature a minimal total defect density is observed for an optimal quench speed. The thermal defect density also increases with increasing temperature.

\begin{figure}
\includegraphics[width=0.47\textwidth]{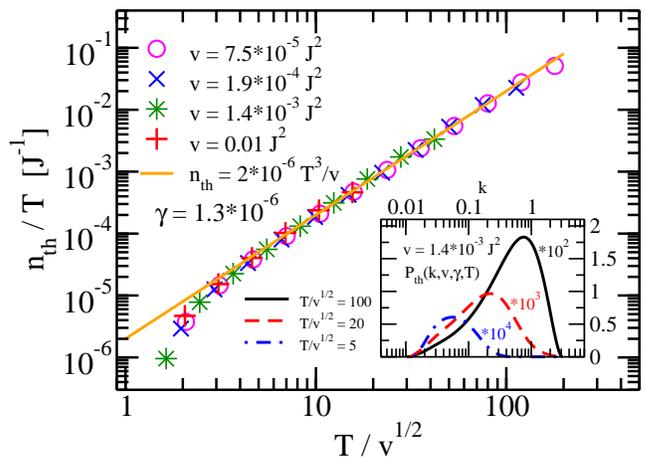}
\caption{(Color online)  Excess thermal defect density associated with the quench, as a function of scaled temperature, for three different quench velocities $v$.  For $\sqrt{v} \lesssim T \lesssim J$, the data follows $n_{\rm th}\propto T^3/v$ in line with the scaling form given in Eq.~(\ref{eq:nth} (orange sold line).
Inset: Plot of momentum-resolved thermal defect density $P_{\rm th}(k,v;\gamma,T)$ for quench velocity $v=1.4\cdot10^{-3}J^2$ and three temperatures.
}
\label{fig:ThermalDefectsScaled}
\end{figure}

To focus on the bath contribution to the defect production during the quench, in Fig.\ \ref{fig:ThermalDefectsScaled} we now plot results for the {\it excess} thermal defect density as a function of scaled temperature.  The analytic estimate predicts a scaling $n_{\rm th}\sim (\kb T)^3/v$ for $\sqrt{v}\ll \kb T \ll J$ (see Eq.~(\ref{eq:nth})) which is indeed observed (see data and orange sold line in Fig.\ \ref{fig:ThermalDefectsScaled}). For smaller $T$, we see that the thermal defect density is suppressed, also in accordance with our prediction that thermal defect production is suppressed in the ``non-equilibrium" regime
$1 / \Delta(t) > |t|$. 

The inset of Fig.\ \ref{fig:ThermalDefectsScaled} plots the momentum-resolved thermal defect density $P_{\rm th}(k,v;\gamma,T)$ exhibiting a clear peak. With increasing temperature the momentum of the peak maximum increases in line with expectation from the behaviour of the minimum in the adiabatic regime of the dissipative Landau-Zener dynamics \citep{NalbachLZ09,NalbachLZ10}. For temperatures $T\rightarrow\sqrt{v}$ the main contribution is from modes at $k\simeq\sqrt{v}$. Note that here the adiabatic - Markovian master equation is least reliable \cite{NalbachLZ14} and thus the observed behaviour in the low temperature regime in Fig.\ \ref{fig:ThermalDefectsScaled} cannot be taken to be conclusive.

%%%%%%%%%%%%%%%%%%%%%%%%%%%%%%%%%%%%%%%%
%%%%%%%%%%%%%%%%%%%%%%%%%%%%%%%%%%%%%%%%

%%%%%%%%%%%%%%%%%%%%%%%%%%%%%%%%%%%%%%%%%%%

\textit{Finite bath correlation length-- }
While the above analysis was for noise that is correlated over the entire lattice, it remains valid for situations where the
bath noise has a {\it finite} spatial correlation length $\xiN$, as long as this length is much longer than all other relevant length scales.
In particular, this requires $\xiN$ be larger than both the Kibble-Zurek length
$\xi_{\rm KZ} = J a / \sqrt{v}$ and the thermal length $\xi_T = J a / k_B T$.  It {\it does not} however require that $\xiN$ approach the size of the system.  The fact that the system correlation length diverges at the QCP is not relevant; heuristically, the bath only plays a role outside the non-equilibrium regime surrounding the QCP, and thus does not see the full divergence of this length.  
A more rigorous justification for why our results apply whenever $\xiN \gg \xi_{\rm KZ}, \xi_T$ (based on a Keldysh analysis) is provided in the EPAPS \cite{suppInfo}.
In the opposite limit of a very small correlation length $\xiN \ll \xi_T$, one approaches the limit of spatially-uncorrelated dissipation, as studied by Patan\`e et al \cite{Patane2008,Patane2009}, with a very different temperature dependence of the thermal defect density from Eq.~(\ref{eq:nth}).  
We thus see that the ratios $\xiN / \xi_{\rm KZ}$ and $\xiN / \xi_{T}$ will in general play a crucial role in determining the influence of dissipation on a quench.

%%%%%%%%%%%%%%%%%%%%%%%%%%%
{\it Conclusions-- }  We have studied the general problem of how thermal dissipation and noise give rise to defect generation in a quantum quench beyond the amount predicted for zero dissipation by the Kibble-Zurek scenario.  We have  explicitly analyzed  quenches in the TFIM coupled to a global thermal bath by mapping to a set of effectively independent dissipative Landau-Zener problems. By analytical estimates as well as numerical calculations, we have shown that the excess thermal defect density scales as $(\kb T)^3/v$ as long as $\sqrt{v}\ll \kb T \ll J$ in clear contrast to previous results obtained for spatially-uncorrelated dissipation. These first steps toward understanding the interplay between quench and dissipation elicit a host of further studies, for instance,  regarding more complex systems, different crossover regimes, general scaling laws, and effect of alternative bath or coupling forms. As with non-dissipative quantum quenches, certain cold atomic and magnetic systems would form the experimental testbeds for these studies. 

We thank Alessandro Silva and Anatoli Polovnikov for useful conversations.
PN acknowledges financial support by the excellence cluster ``The Hamburg Centre for Ultrafast Imaging - Structure, Dynamics and Control of Matter at the Atomic Scale" and by the project ``Nonequilibrium Quantum Energy Transport through Nanostructures" (NA 394/2-1) of the Deutsche Forschungsgemeinschaft.  SV acknowledges the National Science Foundation grant DMR 0644022-CAR.  AAC thanks NSERC. AAC and SV also thank the Kavli Institute for Theoretical Physics for their hospitality.

\bibliographystyle{apsrev}
\bibliography{DissQuenchBib}

\section{Appendix -- Scaling arguments}
Here we generalize the scaling arguments presented in previous works~\cite{PolkovnikovRMP, Chandran2012, Dziarmaga2010} with regards to the universal non-equilibrium dynamcs associated with Kibble-Zurek-type physics.  
We focus on the influence of dissipation from a weakly coupled thermal bath which produces spatially-correlated noise.  We also focus on a system-bath coupling which is analogous to transverse-field noise in the transverse-field Ising model, i.e. the system bath coupling Hamiltonian commutes with the coherent system Hamiltonian away from the critical point.

As the quench process evolves through the quantum critical point, we expect excitations produced by coupling to the thermal bath, $n_{th}$,  to become important in the vicinity of the critical point where the system's energy gap $\Delta$ is small compared to the bath temperature scale $k_BT$. However, very close to the critical point, the system's relaxational time scale, $\tau$, diverges and the quench process becomes too rapid for the system to form the coherences needed to allow bath-induced transitions.  These constraints provide the following bounds for the regime in which thermal excitations are significant and the following scaling forms for their behavior in terms of characteristic energy/time scales.

 Consider a quantum phase transition characterized by a parameter $\alpha$ such that the quantum critical point occurs at a value $\alpha_c$~\cite{Sachdev}. Close to the critical point, the system's typical correlation length scale diverges as
 \begin{equation}
 \xi\sim \delta^{-\nu},
 \label{eq:xi}
 \end{equation}
 where $\nu$ is the associated critical exponent and $\delta=|\alpha-\alpha_c|$ measures the deviation from the critical point. The relaxational time diverges as
 \begin{equation}
 	\tau	\sim \Delta^{-1}\sim\delta^{-\nu z},
 \label{eq:xiT}
 \end{equation}
 where $z$ is the dynamic critical exponent. 

  With regards to  Kibble-Zurek scaling behavior\cite{PolkovnikovRMP, Chandran2012, Dziarmaga2010}, consider a linear quench at a characteristic rate $v^{-1}$,
\begin{equation}
\alpha(t) = \alpha_c+vt.
\label{eq:quench2}
\end{equation}
A given deviation $\delta$ thus occurs at a time $t(\delta) = \delta /v$;
$|t(\delta)|$ represents the time remaining until the system reaches the critical point. Now the quench enters the non-equilibrium regime under the condition 
\begin{equation}
t(\delta) < \tau,
\label{eq:noneq}
\end{equation}
or, from Eq.~(\ref{eq:xiT}), $\delta/v < \delta^{-\nu z}$. This cross-over criterion establishes a scaling relationship between the quench rate and distance to criticality, namely
\begin{equation}
 \delta^{1+\nu z}\sim v.
 \label{eq:deltatau}
 \end{equation}
 From this relationship, one can derive the scaling of standard Kibble-Zurek variables. For instance, the density of defects produced in the non-equilibrium region scales as $n_D\sim\xi^{-d}\sim v^{\nu d/(1+\nu z)}$. In the case of the one-dimensional transverse Ising system, where $\nu=z=d=1$, one obtains the well known scaling behavior $n_D\sim \sqrt{v}$. 

With regards to the regime of interest here, namely, that of thermal excitations, the two constraints mentioned above need to be satisfied. First, as discussed previously in Ref.~\cite{Patane2008, Patane2009},  the bath temperature must be greater than the gap, i.e. $k_B T > \Delta$, yielding the scaling relationship $k_B T \sim \delta^{\nu z}$. Second, the quench must fall short of entering the non-equilibrium regime, or equivalently, $t(\delta) > (k_BT)^{-1}$. This condition, combined with the scaling relationship obtained from the first constraint and with Eq.~(\ref{eq:deltatau}) provides the following temperature lower bound for a given quench rate: 
\begin{equation}
(k_BT)^{(1+\nu z)/(\nu z)}>v,
\label{eq:Ttau}
\end{equation}
as presented in Eq.~(1) in our main text.

In Ref.~\cite{Chandran2012}, similar to equilibrium quantum critical scaling, the effect of finite temperature has been discussed in the context of {\it non-equilibrium} quantum critical scaling. This is done by introducing a dimensionless parameter that is naturally defined by the scaling relationship between temperature and quench rate in Eq.~(\ref{eq:Ttau}). 
It is interesting to note that the excess thermal density that we predict in Eq.~(8) of the main text, given that it has dimensions of inverse volume, is consistent with the scaling form hypothesized in Ref.~\cite{Chandran2012} for generic situations:
\begin{equation}
	n_{th}\sim (k_BT)^{d/z}{\cal F}[(k_BT)^{(1+\nu z)}v^{-\nu z}],
	\label{eq:GeneralScaling}
\end{equation}
where ${\cal F}$ is a scaling function.  Specifically, in the one-dimensional transverse Ising case, our arguments show that thermal excitations are important in the regime $(k_B T)^2>v$ and that they respect the form $n_{th}~\sim k_BT {\cal F}[k_B T/\sqrt{v}]$. 

We emphasize however that care needs to be taken in applying the above scaling. The situation presented by  Ref.~\cite{Chandran2012} presents a closed system having an initial temperature $k_BT$, and in general, unlike our case, thermal effects need not give rise to separate additional contributions above the zero-temperature Kibble-Zurek contributions. We believe that adherence to the expected form is tied to the weak nature of the bath coupling, as well as the particular choice of an Ohmic bath spectral function.
Considering the effects of more general bath spectral functions and stronger couplings would make for an interesting and challenging study.

%%%%%%%%%%%%%%%%%%%%%%%%%%%%%%%%%%%%%%%%%%%%%%%%%%%%%%%%%%%%%%%%%%%%%%%%%

\section{Appendix -- Mapping to independent dissipative Landau-Zener transitions}

In the main text, we argue that in the limiting case where each site of our TFIM couples to the {\it same} dissipative bath (a ``global" system-bath coupling), we can map our
dissipative quench problem onto a set of independent dissipative Landau-Zener problems.  One might worry that this mapping is only approximate, as it ignores correlations between different fermionic modes induced by the globally-coupled bath.  While such correlations will play a role for some physical observables, they play {\it no} role in determining single-particle properties, such as the quasiparticle occupancies which we focus on.  This is a direct consequence of the fact that the system-bath Hamiltonian of Eq.~(3) in the main text conserves the momentum of the fermionic system.

More formally, the occupancy of a given quasparticle mode with momentum $k$ can be written
\begin{eqnarray}
	P_k(t) \equiv  \langle \hat{c}^\dagger_k(t) \hat{c}_k(t) \rangle \equiv -i G^<_k(t,t),
\end{eqnarray}
where we have introduced the standard lesser Keldysh Green's function associated with this mode \cite{KamenevBook}.  If one treats the coupling to the bath as a perturbation of the coherent (dissipation-free) system, the lesser Green's function appearing above is completely determined by the Keldysh self energies $\Sigma^{\alpha}_{k}(t,t')$ associated with the system-bath interaction; here, the index $\alpha$ can take the values $R, A, K$, corresponding to retarded, advanced and Keldysh self-energies \cite{KamenevBook}.  Note the self-energy must be diagonal in momentum, as electronic momentum is conserved.  Consider an arbitrary self energy diagram for $\Sigma^\alpha_{k}(t,t')$.  As fermion momentum is conserved at each system-bath interaction vertex, {\it all} internal fermion propagators in this diagram involve the same momentum $k$.  Heuristically, this implies that at least for single-particle Green's functions, a given fermion mode with momentum $k$ does not know about other modes having a different momentum $k' \neq k$. 

 It follows that we would obtain exactly the same diagrammatic expansion (and hence result for $G^<_k(t,t)$) if we had coupled each fermion mode to its own independent bath.  Formally, this means modifying the system-bath Hamiltonian in Eq.~(3) of the main text as follows:
\begin{eqnarray}
	 \hH_{SB} & \rightarrow &
	 	\sum_{0 \le k \le \pi} \left[\begin{array}{cc}
			\hc^\dagger_{k} & \hc_{-k}
		\end{array} \right]
%		\left( \mathcal{H}_{k,S} + \mathcal{H}_{k,SB} \right)
		\left[\begin{array}{cc}
			\hat{X_k} & 0 \\ 0 & -\hat{X_k}  \end{array} \right]
	\left[\begin{array}{c}
			\hc_{k}
			\\ \hc^\dagger_{-k}
	\end{array} \right] ,
	\nonumber \\
	\label{eq:Hamf2}		
\end{eqnarray}
We now have an independent bath for each $k$ mode, with a corresponding noise operator $\hat{X}_k = \sum_\nu \lambda_{k,\nu} (\hat{b}_{k,\nu} + h.c.) $.  Each of these baths (labelled by $k$) has identical properties to the bath appearing in our starting Hamiltonian.  They all have the same temperature $T$ and identical spectral densities:  $\mathcal{J}_k(\omega)= \mathcal{J}(\omega)$, where $\mathcal{J}(\omega)$ is the spectral density of our original bath (noise operator $\hat{X}$), as given after Eq.~(2) in the main text.  

Thus, for computing quasiparticle occupancies, we can exactly treat each fermion $k$ mode as being effectively coupled to its own independent bath.  This then rigorously justifies our mapping to an ensemble of uncoupled dissipative Landau-Zener problems.  

%%%%%%%%%%%%%%%%%%%%%%%%%%%%%%%%%%%%%%%%%%%%%%%%%%%%%%%%%%%%%%%%%%%%%%%%%

\section{Appendix -- Adiabatic Markovian master equation}

Letting $|j(t)\rangle$ ($j=1,2$) denote the eigenstates of the instantaneous coherent Hamiltonian $\mathcal{H}_{k,S}(t)$ in Eq.~(3) (in the main text) (with eigenergies
$\pm E_k(t)= \pm \sqrt{ \left(\xi_k(t)\right)^2 +\Delta_k^2}\,\,$), the matrix elements of the effective statistical operator $\hat{\rho}_k(t)$ of the Landau-Zener system after tracing out the bath degrees of freedom are parametrized in terms of a 3-vector $\vec{r}_k(t)$ as
$\langle j(t) | \hat{\rho}_k(t) | j'(t) \rangle = \halb(\bbbone - \vec{r}_k \cdot \vec{\tau})_{jj'}$, where $\vec{\tau}$ is the vector of Pauli matrices.
Suppressing the $k$ index for clarity, the adiabatic Markovian master equation takes the form \cite{NalbachLZ14}:
\begin{align}
	\partial_t \vec{r} & =
		\left(\begin{array}{ccc}
		-\gamma_1(t) & 0 & \dot{\theta}(t) \\
		0 & -\gamma_2(t)	&	 - 2E(t)    \\
		-\dot{\theta}(t) & 2E(t)	&	- \gamma_2(t)
	\end{array} \right) \vec{r}
	+  \gamma_1(t)\vec{r}_{\rm eq},
	\label{eq:MasterEqn}
%		\left(\begin{array}{c}
%			0 \\
%			0 \\
%			\tanh[\beta E(t)]
%	\end{array} \right),
\end{align}
where $\theta(t)=\arctan[\xi_k(t) / \Delta_k]$ and $\vec{r}_{\rm eq} = \left(\tanh[\beta E(t)],0,0 \right)$. The terms proportional to $\dot{\theta}$ describe coherent non-adiabatic evolution, while the time-dependent relaxation and dephasing rates are
\bee
	\label{gamrel}
		\gamma_1(t) &=&    \cos^2 \left(\theta(t) \right)  \bar{S}[2 E(t)] \\
	\label{gamdec}
		\gamma_2(t) &=& \halb\gamma_1(t) \,+ \sin^2 \left(\theta(t) \right) \bar{S}[0]
%		\gamma \kb T.
\eee
where $\bar{S}[\omega] =2 \pi \mathcal{J}[\omega] \coth \beta \omega/2$ is the symmetrized spectral density of the bath noise.

For a given quench protocol, i.e. $\xi_k(t)$, we solve the adiabatic-Markovian master equation (\ref{eq:MasterEqn}) using a standard fourth order Runge Kutta scheme \cite{NumRec}. Thus we get the time-dependent statistical operator for each $k$ and, in turn, the probability $P(k,v;\gamma,T)$ for a given effective TLS to end up in the excited state at the end of the quench. To obtain the total excitation density produced during the quench, we integrate over all $k$ using a standard Romberg scheme \cite{NumRec}.

%%%%%%%%%%%%%%%%%%%%%%%%%%%%%%%%%%%%%%%%%%%%%%%%%%%%%%%%%%%%%%%
\section{Appendix -- Finite bath spatial correlations}

As discussed in the main text, our results for a spatially-uniform dissipative bath remain valid in the case where the bath noise has a finite spatial correlation length $\xi_N$, as long as this length is much larger than the Kibble-Zurek length $\xi_{\rm KZ}$; one does not need $\xi_N$ to approach the size of the system.  To make this precise, we generalize the system-bath coupling so that there is a distinct bath noise operator on each site, $\hat{H}_{\rm SB} = \sum_j \hat{\sigma}^z_j \hat{X}_j$.  As usual, we take the bath to be an infinite collection of harmonic oscillators in thermal equilibrium, and take the $\hat{X}_{j}$ to be linear in the bath creation and destruction operators.  We also take the bath to be in a translationally invariant state (unlike the model presented in Ref.~\onlinecite{Patane2009}).  We consider a generic situation where (like the main text) the on-site noise is still described by an Ohmic spectral density $\mathcal{J}[\omega]$,  
but where the noise correlation decays exponentially with distance
\begin{equation}
	\frac{\langle \{ \hat{X}_j[\omega] , \hat{X}_k[\omega'] \} \rangle}{2 \pi \delta(\omega + \omega') } =  
		\mathcal{J}[\omega] \coth\left(\frac{\omega}{ 2 k_B T} \right) e^{-|j-k|/\xi_N}
\end{equation} 

The above form implies that we can express the Fourier-transformed bath noise operators as:
\begin{align}
	\hat{X}_q \equiv \frac{1}{\sqrt{N}} \sum_l e^{-i q l} \hat{X}_l = 
		\Lambda_q \sum_\nu \lambda_\nu  \left( \hat{b}_{q,\nu} + \hat{b}_{-q,\nu}^\dagger \right)
\end{align}
 where the $\hat{b}_{q,\nu}$ describe Einstein phonons with energy $\omega_{\nu}[q] = \omega_\nu$.  We use a normalization such $(1/N) \sum_q |\Lambda_q|^2 = 1$.  The spectral function 
 $\mathcal{J}[\omega]$ associated with the noise on any given site is then identical to that used in the main text.
 
We next specialize to exponentially decaying spatial correlations, with a correlation length much smaller than the system size.  This implies: 
\begin{equation}
	|\Lambda_q|^2 = 
		\frac{ 2 \xi_N / a }{1 + q^2 \xi_N^2} 
\end{equation}

Unlike the global-bath model in the main text, the finite correlation length here means that the bath can exchange non-zero momentum with the system.  We follow 
Ref.~\onlinecite{Patane2009}, and use the weakness of the system-bath coupling to treat the system perturbatively, using a self-consistent Born approximation for the Keldysh self-energy of the fermion Green functions.  
Within this approximation, the Keldysh self energy for a fermion with momentum $k$ is given self-consistently by
\begin{align}
	\label{Eq:SCBSigma}
	& \Sigma_k(t_2,\sigma_2; ,t_1,\sigma_1)  =  \\ 
		& \frac{1}{N} \sum_q \left| \Lambda_q \right |^2
		G_{k+q}(t_2,\sigma_2; ,t_1,\sigma_1)   D_{q}^0(t_2,\sigma_2; ,t_1,\sigma_1),
		\nonumber
\end{align}
where $\sigma_j = \pm$ denotes the forward and backward Keldysh contours, $G_{k}[t,\sigma; t'\sigma']$ is a dressed fermion Keldysh Green function, and $D^0_{q}[t,\sigma; t'\sigma']$ is the unperturbed bosonic (equilibrium) Keldysh Green function for the bath operator $\hat{X}_q / \Lambda_q$.  In our model (where we assume Einstein phonons, corresponding to a frequency-independent $\xi_N$), this Green function is independent of $q$.  

We are now in a position to make estimates concerning the role of $\xi_N$, based on the behaviour of the imaginary part of the self-energies (which control bath-induced transitions).  We will focus on transitions which are thermally enhanced, i.e.~which involve the absorption or emission of bath phonons having $\omega < k_B T$.  For a bath-induced scattering event taking a quasiparticle from momentum $k$ to $k+q$, energy conservation and the fermion dispersion relation will determine the energy of the bath phonon involved.  Assuming $k_B T \ll J$ as always, this then naturally leads to the thermal length $\xi_{T} \equiv a J / k_B T$:  the only transitions that are thermally enhanced involve momentum transfers with  $|q| \lesssim 1 / \xi_T$.  

The simplest regime is where $\xi_N \ll \xi_T$.    In this case, the only thermally-enhanced transitions have $q \ll 1 / \xi_T \ll 1 / \xi_N$, and the $q$-dependence of the structure factor $\Lambda_q$ plays no role:  we can safely replace $\Lambda_q$ by $\Lambda_{q=0}$.  In this case, the $q$ integral in the self energy
of Eq.~(\ref{Eq:SCBSigma}) can be estimated as:  
\begin{align}
	\frac{a}{2 \pi} \int_{-\pi/a}^{\pi/a} |\Lambda_{q}|^2 G_{k+q} & \simeq
		  \frac{a}{2 \pi} |  \Lambda_{0}  |^2 \int_{-1/\xi_T}^{1/\xi_T}   G_{k+q}  \\
	& \propto (\xi_N/a) T
	\label{Eq:LocalEstimate}
\end{align}		
In this limit where $\Lambda_q$ can be treated as a constant, we recover the local-bath model studied in Ref.~\onlinecite{Patane2009}, where there
are no spatial correlations between bath noise operators $\hat{X}_j$.  This mapping to uncorrelated noise is valid irrespective of the value of 
$\xi_{\rm N} / \xi_{\rm KZ}$, where  $\xi_{\rm KZ}= a J/  \sqrt{v}$ is the Kibble-Zurek length introduced in the main text.  Note the explicit factor of $T$ that emerges from the momentum summation.  

Consider next the opposite regime, where $\xi_N \gg \xi_T$.
In this case, the structure factor $\Lambda_q$ will suppress the contribution of large momentum transfers in the self-energy, as opposed to the bath temperature:
the largest contributing $|q|$ will be $\sim 1 / \xi_N$.  If in addition we have $\xi_N \gg \xi_{\rm KZ}$, then we can also ignore the $q$ dependence of the fermion propagator $G_{k+q}$ in Eq.~(\ref{Eq:SCBSigma}).  To understand this point, note that without dissipation, quasiparticle modes with $k \ll 1 / \xi_{\rm KZ}$ will evolve diabatically during the quench (and become excited), while modes with $k \gg 1 / \xi_{\rm KZ}$ will evolve adiabatically (and hence remain unpopulated).
Correspondingly, for small $k$, the quasiparticle modes $k$ and $k+q$ behave almost identically when $|q| \ll 1 / \xi_{\rm KZ}$, implying $G_{k} \sim G_{k+q}$.

Thus, for $\xi_N \gg \xi_{\rm KZ}, \xi_T$, the $q$ summation in the self-energy of Eq.~(\ref{Eq:SCBSigma})  can be estimated as 
\begin{align}
	\frac{a}{2 \pi} \int_{-\pi/a}^{\pi/a} |\Lambda_{q}|^2 G_{k+q} & \simeq
	\frac{a}{2 \pi} G_k \int_{-\pi/a}^{\pi/a} |\Lambda_{q}|^2  \\
	 & \simeq
		G_{k}  \frac{a}{2 \pi} \int_{-\infty}^{\infty} |\Lambda_{q}|^2  
		\simeq G_{k} 
\end{align}
This is identical to having taken the global-coupling, $\xi_N \rightarrow \infty$ limit from the outset, i.e.~having used
\begin{equation}
	\Lambda_q = \frac{2 \pi}{a} \delta(q)
\end{equation}
Thus, when $\xi_N$ is the largest length scale in the problem (i.e.~$\xi_N \gg \xi_{\rm KZ}, \xi_{\rm T}$), the system does not know about the finite bath spatial correlation length $\xi_N$, and one gets the same results as for a model where $\xi_N \rightarrow \infty$.   Note that in this large $\xi_N$ limit, the summation over $q$ gives a temperature-independent result, in contrast to the small $\xi_N$ estimate in 
Eq.~(\ref{Eq:LocalEstimate}).  This difference is at the heart of why our global-coupling result for the thermal defect density scales as a lower power of temperature than the corresponding result found in Refs.~\onlinecite{Patane2008,Patane2009} for a locally-coupled bath.

Finally, the is the remaining case $\xi_{T} \ll \xi_N \ll \xi_{\rm KZ}$.  In this case, the bath correlation length cuts off large momentum transfers (as opposed to temperature).    However, this cutoff is not large enough to prevent coupling between very different fermionic modes, i.e.~adiabatic and non-adiabatic modes.  In this cross-over regime, neither the local model studied in 
Ref.~\onlinecite{Patane2008} nor global bath model studied in the main text are appropriate.

\end{document}